\documentstyle[preprint,prc,eqsecnum,aps,epsf]{revtex}
\begin{document}
\draft

\title{Elastic $e$-$d$ Scattering Data and the Deuteron Wave Function}
\author{R.\ Schiavilla}
\address{Jefferson Lab, Newport News, Virginia 23606 \\
         and \\
         Department of Physics, Old Dominion University, Norfolk, Virginia 23529}
\author{V.R.\ Pandharipande}
\address{Department of Physics, University of Illinois at Urbana-Champaign, Urbana, IL 61801}
\date{\today}
\maketitle

\begin{abstract}
What range of momentum components in the deuteron
wave function are available $e$$d$ elastic scattering
data sensitive to ?  This question is addressed within
the context of a model calculation of the deuteron form
factors, based on realistic interactions and currents.
It is shown that the data on the $A(q)$, $B(q)$, and
$T_{20}(q)$ observables at $q \leq 6$ fm$^{-1}$ essentially
probe momentum components up to $\simeq 4\, m_\pi$.
\end{abstract}

\section{Introduction, Results, and Conclusions}
\label{sec:intro}

\begin{figure}[bth]
\let\picnaturalsize=N
\def\picsize{4in}
\def\picfilenamea{deut_jun.eps}
\ifx\nopictures Y\else{\ifx\epsfloaded Y\else\input epsf \fi
\let\epsfloaded=Y
\centerline{
\ifx\picnaturalsize N\epsfxsize \picsize\fi \epsfbox{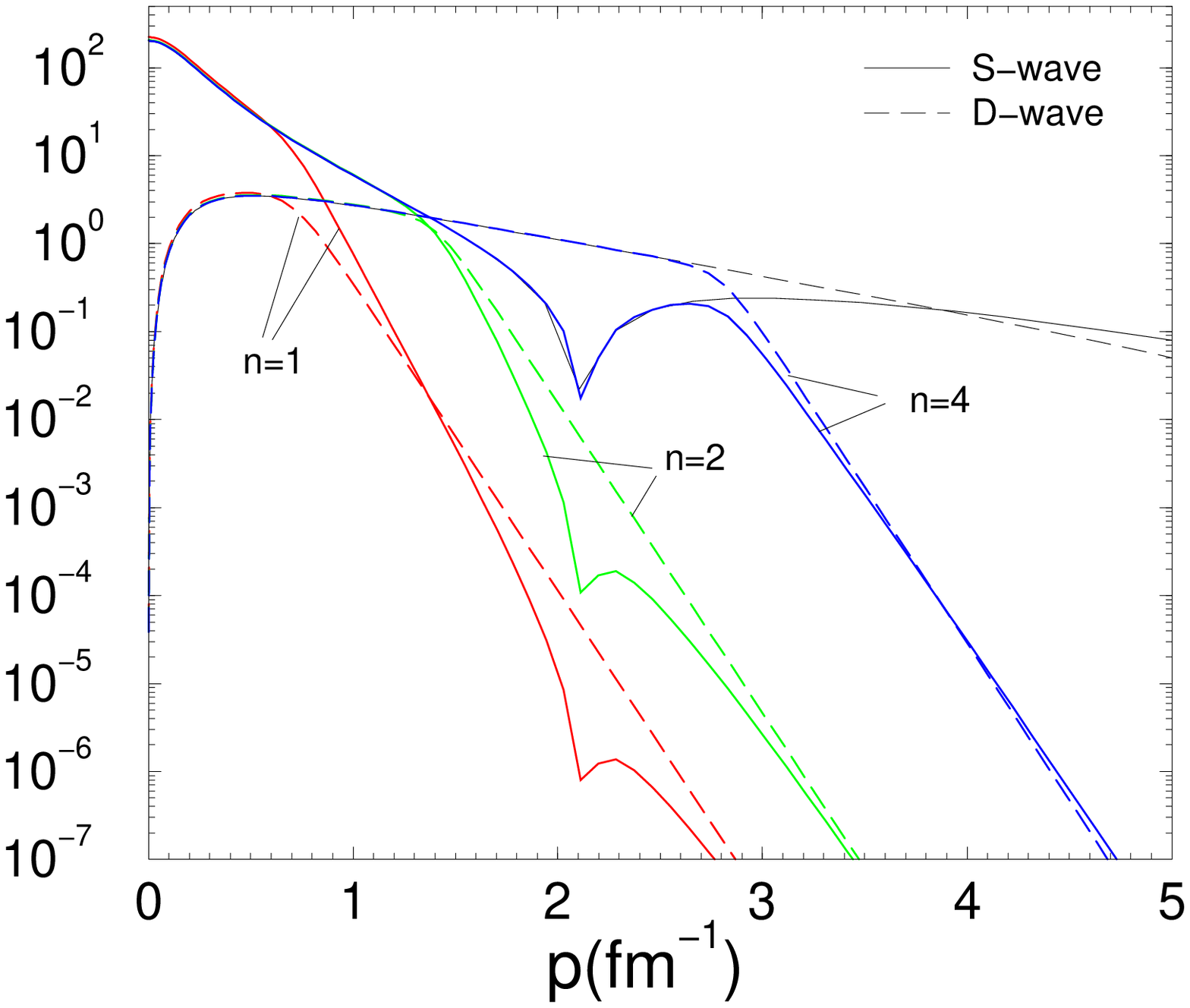}
 }}\fi
\caption{The truncated wave functions $\overline{u}_L(p;n)$, with
$n$=1, 2 and 4, are compared with the reference $u_L(p)$.}
\label{fig:deut}
\end{figure}

\begin{figure}[bth]
\let\picnaturalsize=N
\def\picsize{4in}
\def\picfilenamea{a_jun.eps}
\ifx\nopictures Y\else{\ifx\epsfloaded Y\else\input epsf \fi
\let\epsfloaded=Y
\centerline{
\ifx\picnaturalsize N\epsfxsize \picsize\fi \epsfbox{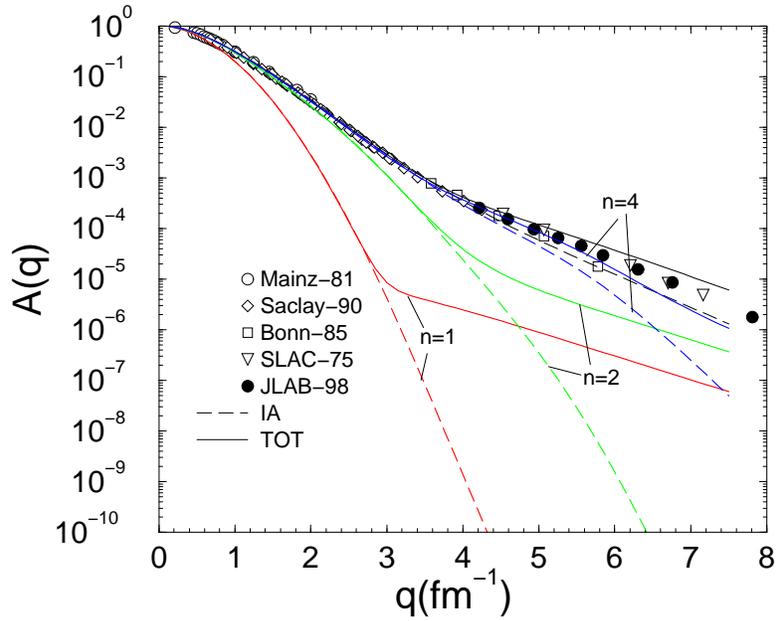}
 }}\fi
\caption{The $A(q)$ structure functions obtained with the
truncated wave functions $\overline{u}_L(p;n)$ for $n$=1,2, and 4
are compared with the reference $A(q)$ and data.  The results
obtained with one-body only and both two- and two-body 
operators, labelled respectively IA and TOT, are shown.}
\label{fig:a}
\end{figure}

\begin{figure}[bth]
\let\picnaturalsize=N
\def\picsize{4in}
\def\picfilenamea{b_jun.eps}
\ifx\nopictures Y\else{\ifx\epsfloaded Y\else\input epsf \fi
\let\epsfloaded=Y
\centerline{
\ifx\picnaturalsize N\epsfxsize \picsize\fi \epsfbox{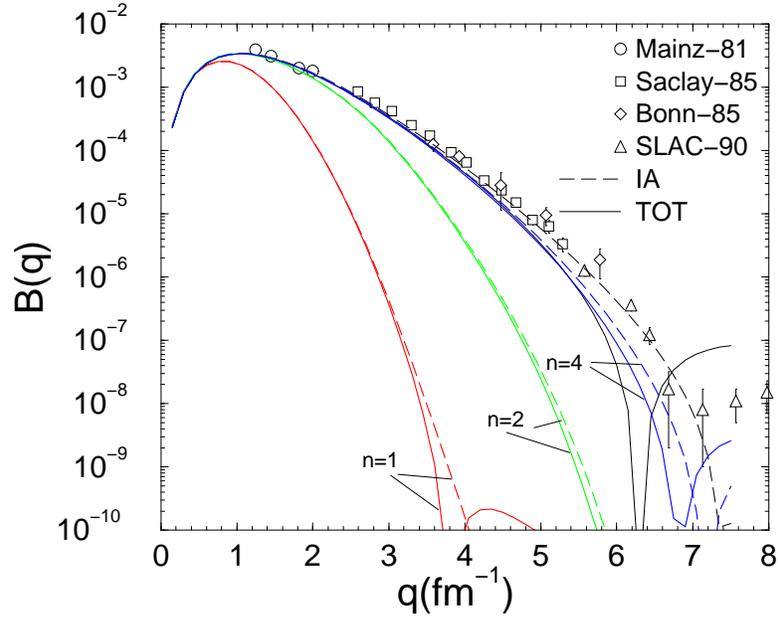}
 }}\fi
\caption{Same as in Fig.~\protect\ref{fig:a}, but for the
$B(q)$ structure function.}
\label{fig:b}
\end{figure}

\begin{figure}[bth]
\let\picnaturalsize=N
\def\picsize{4in}
\def\picfilenamea{t20_jun.eps}
\ifx\nopictures Y\else{\ifx\epsfloaded Y\else\input epsf \fi
\let\epsfloaded=Y
\centerline{
\ifx\picnaturalsize N\epsfxsize \picsize\fi \epsfbox{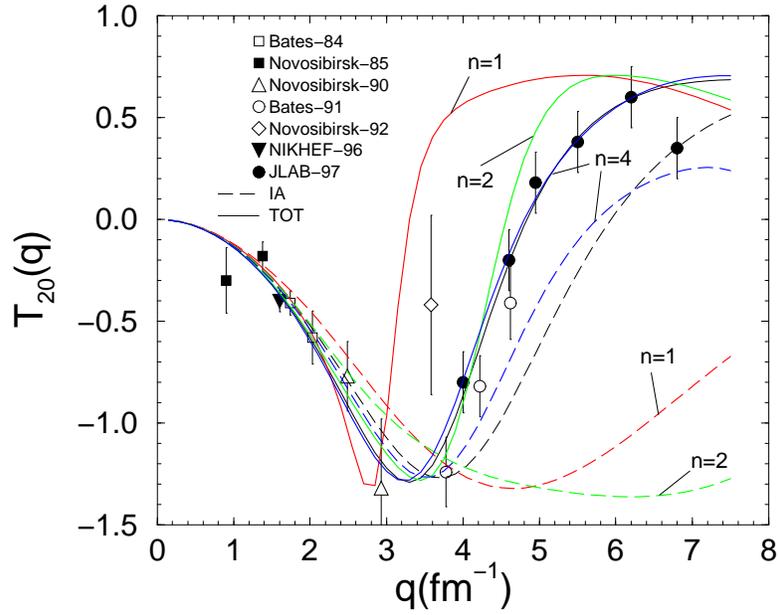}
 }}\fi
\caption{Same as in Fig.~\protect\ref{fig:a}, but for the
$T_{20}(q)$ tensor polarization.}
\label{fig:t20}
\end{figure}

The present work addresses the following issue: what range
of momentum components in the deuteron wave function are probed
by presently available $e$-$d$ elastic scattering \hbox{data ?}
The question is answered within the context of a model~\cite{Schiavilla02}, 
based on realistic interactions
and currents, that predicts quite well the observed deuteron
$A(q)$ and $B(q)$ structure functions and $T_{20}(q)$ tensor
polarization up to momentum transfers $q \simeq 6$ fm$^{-1}$.
To this end, the deuteron S- and D-wave components obtained
in the full theory~\cite{Schiavilla02} and denoted
as $u_L(p)$, $L$=$0, 2$, are truncated, in momentum space, as

\begin{equation}
\overline{u}_L(p;n) =
\frac{c_n}{\left[ 1 + {\rm exp}(p-p_n)/a \right]^{1/2} } \, u_L(p) \>\>,
\end{equation}
where the cutoff momentum $p_n$=$n\, m_\pi$ ($m_\pi$ is the pion mass and $n$ is an
integer) and $a$=$0.1\, m_\pi$.  The constants $c_n$ are fixed through the
normalization condition, Eq.~(\ref{eq:norm}) below.  The wave functions
$\overline{u}_L(p;n)$ corresponding to $n$=1, 2 and 4 along with
the reference $u_L(p)$ are shown in Fig.~\ref{fig:deut}.  The $A(q)$,
$B(q)$, and $T_{20}(q)$ observables are calculated using these
truncated wave functions by including one-body only and both
one- and two-body currents.  The results are displayed in
Figs.~\ref{fig:a}--\ref{fig:t20}.  

Aspects of the present theory~\cite{Schiavilla02} pertaining
to the input potential, the boost corrections in the wave
functions, the one and two-body charge and current operators,
are succinctly summarized in Secs.~\ref{sec:wave}--\ref{sec:ffs}.
Here it suffices to note that the results shown in
Figs.~\ref{fig:a}--\ref{fig:t20} are obtained within a scheme
in which these various facets are consistently
treated to order $(v/c)^2$.  Thus the present approach
improves and extends that adopted in Refs.~\cite{Schiavilla91}
and~\cite{Wiringa95}, in which relativistic corrections
of order $(v/c)^2$ were selectively retained.  For example,
the $(v/c)^2$ contributions associated with the $\pi$-exchange
charge operator were included, while those originating from
the one-pion-exchange potential and from boosting the wave
function were ignored.  The consistent $(v/c)^2$ scheme employed here,
however, does not alter quantitatively the predictions obtained
in Refs.~\cite{Schiavilla91,Wiringa95}, except for the $B(q)$ structure
function as discussed in Sec.~\ref{sec:res}.
A detailed analysis of the differences between the two approaches
is beyond the scope of this work; it will be presented
in Ref.~\cite{Schiavilla02}.

We assume that the three-momentum transfer ${\bf q}$ in $e$-$d$ 
elastic scattering is in the $z$-direction.  In the Breit frame
the deuteron has initial and final momenta $-q/2$ and $+q/2$ in
this direction, respectively.  In the presence of only single-nucleon 
currents, the deuteron wave function has to have components with 
relative momenta $p$=$\mid {\bf p}_1-{\bf p}_2\mid/2$ larger than
$q/4$ in order to produce elastic scattering at momentum 
transfer $q$.  In a limiting case, for example, the nucleons have momenta 
of $-q/2$ and $0$ in the initial state and $+q/2$ and $0$ in the final, 
all along the $z$-axis.  Thus elastic scattering via one-body currents 
probes momentum distributions $u_L(p)$ at $p > q/4$.
This argument, however, does not establish the maximum value of $p$ 
which the observed form factors with $q \leq 6$ fm$^{-1}$
are sensitive to.  In addition, there is no kinematical limit
on initial and final relative momenta for scattering via pair
currents.  Therefore, we use a realistic model of the deuteron
to estimate the maximum value of $p$ probed by the available data. 
 
There is growing interest in chiral effective field theories of nuclear 
forces and currents~\cite{VanKolck99,Walzl01}.  In such theories the effective 
Lagrangian is obtained after integrating out states with momenta greater 
than a specified cutoff.  We hope that the present results provide an 
estimate of the deuteron structure obtained in theories with cutoff 
momentum $p_n$=$n\, m_\pi$. 

The deuteron wave function is dominated by the one-pion exchange 
potential $v_\pi({\bf k})$, where ${\bf k}$=${\bf p}^{\prime} - {\bf p}$ 
is the momentum transferred by the exchanged pion.  The dependence of
$v_\pi({\bf k})$ (see Eq.~(\ref{eq:vpai}) below) on the magnitude $k$
is primarily given by the $\pi NN$ form factor $f_\pi(k)$ for
$k > m_\pi$.  In this limit the leading non-relativistic term of 
$v_\pi({\bf k})$ becomes:

\begin{eqnarray}
v_\pi({\bf k})&=&-\frac{f_{\pi NN}^2}{m_\pi^2}
\frac{f^2_\pi(k)}{ m_\pi^2+k^2}\, {\bbox \sigma}_1 \cdot {\bf k} \,
                                  {\bbox \sigma}_2 \cdot {\bf k} 
\,{\bbox \tau}_1 \cdot {\bbox \tau}_2 \nonumber \\
&\simeq& -\frac{f_{\pi NN}^2}{m_\pi^2}\, f^2_\pi(k)\,
{\bbox \sigma}_1 \cdot \hat{{\bf k}} \,
{\bbox \sigma}_2 \cdot \hat{{\bf k}} 
\,{\bbox \tau}_1 \cdot {\bbox \tau}_2 \ ,
\end{eqnarray}
where $\hat{\bf k}$ is a unit vector.  Thus the deuteron
wave function at relative momentum $p$ is expected to be sensitive to
$f_\pi(k)$ up to $k \simeq p$. 

The results shown in Figs.~\ref{fig:a}--\ref{fig:t20} indicate
that the available data on deuteron form factors at $q \leq 6$
fm$^{-1}$ confirm the present wave functions up to $p \simeq 4\, m_\pi$. 
After including pair-current contributions, the reference wave function 
and that truncated at $p$=$4\, m_{\pi}$ do not give significantly
different form factors in this $q$-range.  The reference and truncated
wave functions, respectively, over- and under-predict the observed $A(q)$
at $q \geq 6$ fm$^{-1}$; both under-predict $B(q < 7$ fm$^{-1})$,
and fail to reproduce the data point for $T_{20}$ at $q \simeq 6.8$ fm$^{-1}$.
The available data on $A(q)$ at larger values of $q$ can be 
used to test the deuteron wave function at $p > 4\, m_{\pi}$; however,
improved theoretical understanding and more accurate data on $B(q>6$ fm$^{-1})$,
and new data on $T_{20}(q > 6$ fm$^{-1})$ are needed.

\section{Deuteron Wave Function}
\label{sec:wave}

The deuteron rest-frame wave function is obtained
by solving the momentum-space Schr\"odinger equation
with the relativistic Hamiltonian~\cite{Forest99,Forest00}

\begin{equation}
H^\mu = 2\sqrt{p^2+m^2} + v^\mu \>\>,
\end{equation}
where $v^\mu$ consists of a short-range part $v_R^\mu$ parameterized
as in the Argonne $v_{18}$ potential~\cite{Wiringa95}, and of a
relativistic one-pion-exchange potential (OPEP) given by

\begin{eqnarray}
v^\mu_\pi ({\bf p}_1^\prime,{\bf p}_1)=&-&\frac{f_{\pi NN}^2}{m_\pi^2}
\frac{f^2_\pi(k)}{ m_\pi^2+k^2}
\frac{m}{E^\prime} \frac{m}{E}
\Bigg[ {\bbox \sigma}_1 \cdot {\bf k} \,
  {\bbox \sigma}_2 \cdot {\bf k} \nonumber \\
&+& \mu \times (E^\prime-E)
\left( \frac{ {\bbox \sigma}_1 \cdot {\bf p}_1^\prime \,
              {\bbox \sigma}_2 \cdot {\bf p}_1^\prime } {E^\prime+m}
      -\frac{ {\bbox \sigma}_1 \cdot {\bf p}_1 \,
              {\bbox \sigma}_2 \cdot {\bf p}_1 } {E+m} \right) \Bigg]
{\bbox \tau}_1 \cdot {\bbox \tau}_2 \>\>.
\label{eq:vpai}
\end{eqnarray}
Here $m$ is the nucleon mass,
$f_{\pi NN}$ is the pion-nucleon coupling constant
($f^2_{\pi NN}/4\pi$=0.075), ${\bf p}_1$ and ${\bf p}_1^\prime$ are
the initial and final momenta of particle 1 in the center-of-mass
frame, ${\bf k}$=${\bf p}_1^\prime-{\bf p}_1$ is the momentum transfer,
$E$=$\sqrt{p_1^2+m^2}$, and $E^\prime$=$\sqrt{p_1^{\prime\, 2}+m^2}$. 
The monopole form factor: 
\begin{equation}
f_\pi(k)=\frac{\Lambda_\pi^2-m_\pi^2}{\Lambda_\pi^2+k^2} \>\>,
\end{equation}
with $\Lambda_\pi$=1.2 GeV/c is used in this work.  

The $\mu$-dependent term
characterizes possible off-energy-shell
extensions of OPEP, and leads to strong non-localities
in configuration space.  In particular, the value $\mu$=--1
($\mu$=1) is predicted by pseudoscalar (pseudovector) coupling
of pions to nucleons, while $\mu$=0 corresponds to the so-called
\lq\lq minimal non-locality\rq\rq choice~\cite{Friar77}.
It has been known for over two decades~\cite{Friar77}, and recently
re-emphasized by Forest~\cite{Forest00} that these various
off-shell extensions of OPEP are related to each other
by a unitary transformation, in the sense that

\begin{equation}
{\rm e}^{-{\rm i}\mu U} H^{\mu=0} {\rm e}^{{\rm i}\mu U} \simeq H^{\mu=0}
+ {\rm i} \mu \Big[ H^{\mu=0} \, , \, U \Big] \simeq H^\mu \>\>,
\end{equation}
if terms of $2\pi$-range (and shorter-range) are neglected.
The hermitian operator $U$ is given explicitly in
Ref.~\cite{Forest00}.  This unitary equivalence implies that predictions for
electromagnetic observables, such as the form factors under
consideration here, are independent of the particular off-shell
extension adopted for OPEP, provided that the electromagnetic
current operator, specifically its two-body components associated
with pion exchange, are constructed consistently with this off-shell
parameter.  This point will be further elaborated below.  At this
stage, however, it is important to recall that Forest has constructed
relativistic Hamiltonians $H^\mu$ with $\mu$=$\pm 1,0$, each
designed to be phase-equivalent to the non-relativistic $H$, based
on the Argonne $v_{18}$ potential.

Given this premise, the $\mu$=0 prescription is adopted
for OPEP from now on, and the superscript is
dropped from $H^{\mu=0}$, for ease of presentation.
The resulting momentum-space wave function in the rest frame
of the deuteron is denoted with $\psi_M({\bf p};0)$ (here,
the argument $0$ indicates the rest frame in which the 
deuteron has velocity ${\bf v}$=0), and is written as

\begin{equation}
\psi_M({\bf p};0) = \left[ u_0(p) {\cal Y}_{011}^M(\hat{\bf p}) +
                           u_2(p) {\cal Y}_{211}^M(\hat{\bf p}) \right] \eta^0_0 \>\>,
\end{equation}
where ${\bf p}$ is the relative momentum, $M$ is the
angular-momentum projection along the $z$-axis,
$\eta^0_0$ is the pair isospin $T$=0 state,
and ${\cal Y}_{LSJ}^M(\hat{\bf p})$ are standard spin-angle functions.
The normalization is given by: 
\begin{equation}
\int_0^\infty \frac{{\rm d}p\, p^2}{(2\pi)^3} [ u_0^2(p)+u_2^2(p) ]=1 \>\>.
\label{eq:norm}
\end{equation}

The internal wave function in a frame 
moving with velocity ${\bf v}$ with respect to the
rest frame can be written to order $(v/c)^2$ as~\cite{Friar77}

\begin{eqnarray}
\psi_M({\bf p};{\bf v})&\simeq&\left( 1-\frac{v^2}{4}\right)
 \left[ 1-\frac{1}{2} ({\bf v} \cdot {\bf p})
({\bf v} \cdot \nabla_p) -\frac{{\rm i}}{4m} {\bf v}
\cdot (\bbox{\sigma}_1-\bbox{\sigma}_2)\times {\bf p} \right] \psi_M({\bf p};0) \nonumber \\
&\simeq& \frac{1}{\sqrt{\gamma}} \left[ 1 -\frac{{\rm i}}{4m} {\bf v}
\cdot (\bbox{\sigma}_1-\bbox{\sigma}_2)\times {\bf p} \right]
\psi_M({\bf p}_\parallel/\gamma, {\bf p}_\perp;0) \>\>,
\label{eq:boost}
\end{eqnarray}
where $\gamma = 1/\sqrt{1-v^2}$, and ${\bf p}_\parallel$
and ${\bf p}_\perp$ denote the components of the momentum
${\bf p}$ parallel and perpendicular to ${\bf v}$, respectively.
Only the kinematical boost corrections, associated with
the Lorentz contraction (term $\propto {\bf v} \cdot {\bf p}
\, {\bf v} \cdot \nabla_p$) and Thomas precession of the
spins (term $\propto {\bf v}\cdot (\bbox{\sigma}_1-\bbox{\sigma}_2)\times {\bf p}$)
are retained in Eq.~(\ref{eq:boost}).  There are in principle additional,
interaction-dependent boost corrections.  Those originating
from the dominant one-pion-exchange component of the
interaction have been constructed explicitly in Ref.~\cite{Friar77}
to order $(v/c)^2$.  At this order, however, their contribution
to the form factors in Eqs.~(\ref{eq:g0})--(\ref{eq:g2}) vanishes,
since it involves matrix elements of
an odd operator under spin exchange between $S$=1 states.
Indeed, this same selection rule also holds 
for the Thomas precession term.  Therefore the interaction-dependent
boost corrections do not contribute to elastic scattering up to order
$(v/c)^2$, and are neglected in the following. 

It should also be noted that terms of order higher than $(v/c)^2$
due to Lorentz contraction have in fact been included
in the second line of Eq.~(\ref{eq:boost}).  The factor
$1/\sqrt{\gamma}$ ensures that the wave function in the moving frame
is normalized to one, ignoring corrections of order
$(v/c)^4$ from the Thomas precession term.  

It is interesting to study the relation between the Breit-frame matrix
element $\overline{\rho}({\bf q}; B)$ of the point-nucleon density operator, 
and the rest frame $\overline{\rho}({\bf q}; 0)$.
One finds, again ignoring Thomas precession contributions: 
 
\begin{eqnarray}
\overline{\rho}_M({\bf q};B)&=&\int \frac{{\rm d}{\bf p}}{(2\pi)^3}
\psi_M^\dagger({\bf p}+{\bf q}/4;{\bf v}_B) \,\,
\psi_M({\bf p}-{\bf q}/4;-{\bf v}_B) \nonumber \\
&=&\frac{1}{\gamma_B} \int \frac{{\rm d}{\bf p}}{(2\pi)^3}
\psi_M^\dagger\left(\frac{{\bf p}_\parallel+{\bf q}/4}{\gamma_B},
{\bf p}_\perp;0\right) \,\,
\psi_M\left(\frac{{\bf p}_\parallel-{\bf q}/4}{\gamma_B},
{\bf p}_\perp;0\right) \nonumber \\
&=&\overline{\rho}_M({\bf q}/\gamma_B;0) \>\>,
\end{eqnarray}
after rescaling of the integration variables,
$({\bf p}_\parallel/\gamma_B,{\bf p}_\perp) \rightarrow
({\bf p}_\parallel,{\bf p}_\perp)$, in the last integral. 
Here $\gamma_B$ is the Lorentz factor
corresponding to ${\bf v}_B =(q/2)
\hat {\bf z}/\sqrt{q^2/4+m_d^2}$ in the Breit frame.
This result is consistent with the naive expectation
that the density in configuration space is \lq\lq squeezed\rq\rq~in
the direction of motion by the
Lorentz factor $\gamma_B$ or, equivalently, that its
Fourier transform is \lq\lq pushed out\rq\rq~by $\gamma_B$.
For $q$=6 fm$^{-1}$, $v_B \simeq 0.3$ and $\gamma_B \simeq 1.05$,
corresponding to a 5 \% Lorentz contraction.

\section{Nuclear Electromagnetic Current}
\label{sec:current}

The electromagnetic charge and current operators include
one- and two-body terms.  Only the isoscalar parts
need to be considered, since the deuteron has $T$=0. 
The one-body term is taken as

\begin{equation}
j_{\rm 1-body}^\sigma({\bf q})=
(\rho_{\rm 1-body}({\bf q}),{\bf j}_{\rm 1-body}({\bf q}))=
\sum_{i=1,2} \frac{1}{2} \overline{u}_i^\prime
\left[ F^S_1(q) \gamma^\sigma + 
{\rm i}\frac{F^S_2(q)}{2m} \sigma^{\sigma \tau} q_\tau \right] u_i \>\>,
\end{equation}
where $q^\sigma$=$(0, q \hat{\bf z})$ in the Breit frame, $u_i$ and
$u_i^\prime$ are the initial and final spinors of nucleon
$i$, respectively, with $\overline{u}_i \equiv u_i^\dagger \gamma^0$,
and $F_1^S$ and $F_2^S$ denote the isoscalar combinations of the nucleon's
Dirac and Pauli form factors, normalized as $F^S_1(0)$=1
and $F^S_2(0)$=$-0.12$ (in units of n.m.).  The H\"ohler
parameterization~\cite{Hohler76} of $F_1$ and $F_2$ is used in this 
work.  The spinor $u$, or rather its transpose, is given by

\begin{equation}
u^T=\left( \frac{E+m}{2 E}\right)^{1/2}
\left( \chi_s\, , \, \frac{ {\bbox \sigma} \cdot
 {\bf p} }{E+m} \chi_s\right) \>\>,
\end{equation}
where ${\bf p}$ and $E$=$\sqrt{p^2+m^2}$ are
the nucleon's momentum and energy, and $\chi_s$ is its
(two-component) spin state.  Note that $u^\dagger u$=$1$.
In earlier published work on the form factors
of the deuteron~\cite{Schiavilla91,Wiringa95}
and A=3--6 nuclei, most recently~\cite{Marcucci98,Wiringa98}, boost
contributions were neglected, and only
terms up to order $(v/c)^2$ were included in the non-relativistic expansion
of $j_{\rm 1-body}^\sigma$, namely the well known Darwin-Foldy
and spin-orbit corrections to $\rho({\bf q})$.  In the calculations
reported here, however, the full Lorentz structure
of $j_{\rm 1-body}^\sigma$ is retained.

The two-body terms included in $j_{\rm 2-body}^\sigma$ are those associated 
with $\pi$- and $\rho$-meson exchanges and the $\rho\pi\gamma$
transition mechanism.  The $\pi$ and $\rho$ spatial current operators
to leading order in $v/c$ are isovector, and therefore their
contributions vanish in $e$-$d$ elastic scattering.  The $\pi$-exchange
charge operator is obtained, consistently
with the off-shell parameter $\mu$ adopted for OPEP, as~\cite{Friar77}

\begin{equation}
\rho_\pi^\mu({\bf q})=\frac{(3-\mu)}{8m}\frac{f_{\pi NN}^2}{m_\pi^2}
F_1^S(q) \frac{ f_\pi(k_2) }{k_2^2+m_\pi^2} {\bbox \sigma}_1 \cdot {\bf q}
\,\, {\bbox \sigma}_2 \cdot {\bf k}_2 \, {\bbox \tau}_1 \cdot {\bbox \tau}_2
+ 1 \rightleftharpoons 2 \>\>,
\end{equation}
where ${\bf k}_i$=${\bf p}_i^\prime-{\bf p}_i$,
$f_\pi(k)$ is the $\pi$$N$$N$ form factor defined in Sec.~\ref{sec:wave},
and $F_1^S(q)$ denotes the H\"ohler parameterization~\cite{Hohler76}
for the nucleon's isoscalar Dirac form factor.  It is easy to show,
to order $(v/c)^2$, that

\begin{equation}
{\rm e}^{-{\rm i}\mu U} \left( \rho_{\rm 1-body}
+\rho_\pi^{\mu=0} \right) {\rm e}^{{\rm i}\mu U} \simeq \rho_{\rm 1-body}
+\rho_\pi^{\mu=0}
+ {\rm i} \mu \Big[ \rho_{\rm 1-body} \, , \, U \Big] \simeq 
\rho_{\rm 1-body}+\rho_\pi^\mu \>\>,
\end{equation}
where exp($-{\rm i}\mu U$) is the unitary transformation of Sec.~\ref{sec:wave}.  Again,
we emphasize that the choice $\mu$=0 is made for $\rho_\pi^\mu$ (as for OPEP
in Sec.~\ref{sec:wave}) in the present work.

At this point it is important to recall that
the $\pi$-exchange charge operator corresponding
to $\mu$=--1 was included in all our previous
studies of light nuclei form factors
(for a review, see Ref.~\cite{Carlson98}), based on
non-relativistic Hamiltonians, in which the OPEP
part of $v$ was taken to be its leading local form, i.e.
$E^\prime$=$E$ and $E/m \rightarrow 1$ in Eq.~(\ref{eq:vpai}).
While these calculations are not strictly consistent, since
relativistic corrections are selectively retained in $\rho({\bf q})$ 
but ignored in the potentials and wave functions, they nevertheless
give fairly accurate results since the corrections to 
the wave function are rather small \cite{Forest99}. 

Vector-meson ($\rho$ and $\omega$) two-body charge operators
have been found to give contributions, in calculations
of light nuclei form factors~\cite{Carlson98}, that are typically
an order of magnitude smaller than those associated
with the $\rho_\pi$ operator.  In the present study,
only the $\rho$-exchange charge operator is considered:

\begin{equation}
\rho_\rho({\bf q})=\frac{g_{\rho NN}^2 ( 1 +\kappa_{\rho NN} )^2}{8 m^3}
F_1^S(q) \frac{ f_\rho (k_2) }{k_2^2+m_\rho^2}
\left({\bbox \sigma}_1 \times {\bf q}\right)
\cdot \left( {\bbox \sigma}_2 \times {\bf k}_2\right) \, 
{\bbox \tau}_1 \cdot {\bbox \tau}_2
+ 1 \rightleftharpoons 2 \>\>,
\end{equation}
where $m_\rho$, $g_{\rho NN}$ and $\kappa_{\rho NN}$ are the
$\rho$-meson mass, vector and tensor coupling constants,
respectively.  Here the values $g^2_{\rho NN} /(4\pi)$=0.84,
and $\kappa_{\rho NN}$=6.1 are used from the Bonn 2000
potential~\cite{Machleidt01}, while the $\rho NN$ monopole
form factor $f_\rho(k)$ has $\Lambda_\rho$=1.2 GeV/c. 

Finally, the $\rho\pi\gamma$ current is obtained from
the associated Feynman amplitude as 

\begin{eqnarray}
j^\mu_{\rho\pi\gamma}({\bf q})&=&{\rm i} 
g_{\rho\pi\gamma} \frac{ g_{\rho NN} }{m_\rho} \frac{f_{\pi NN} }{m_\pi}
G_{\rho\pi\gamma}(q)\, 
\frac{f_\rho(k_1)}{k_1^2+m_\rho^2}\,
\frac{f_\pi(k_2)}{k_2^2+m_\pi^2}\,
\epsilon^{\mu\nu\sigma\tau} k_{1,\sigma} q_\tau \nonumber \\
&\times&\left[ \overline{u}_1^\prime
\left( \gamma_\nu +{\rm i} \frac{\kappa_{\rho NN} }{2m} 
\sigma_{\nu \alpha} k_1^\alpha \right) u_1\right]
\left( \overline{u}_2^\prime \,
\gamma_\beta k^\beta_2 \, \gamma_5 u_2 \right) \,
{\bbox \tau}_1 \cdot {\bbox \tau}_2
 + 1 \rightleftharpoons 2 \>\>,
\label{eq:rpg}
\end{eqnarray}
where $g_{\rho\pi\gamma}$ and $G_{\rho\pi\gamma}(q)$ are
the $\rho\pi\gamma$ coupling constant and form factor, respectively,
and $\epsilon_{0123}$=1.  The value $g_{\rho\pi\gamma}$=0.56
is obtained from the measured width of the decay
$\rho \rightarrow \pi \gamma$~\cite{Berg80}, while the form
factor is modeled, using vector-meson dominance, as

\begin{equation}
G_{\rho\pi\gamma}(q)=\frac{1}{1+q^2/m^2_\omega} \>\>,
\end{equation}
where $m_\omega$ is the $\omega$-meson mass.  Note that in 
Eq.~(\ref{eq:rpg}) the term proportional to $k_1^\alpha k_1^\beta$
in the $\rho$-meson propagator has been dropped, since it
vanishes if the nucleons are assumed to be on
mass shell.  Furthermore, in both meson propagators
retardation effects have been neglected.

The leading terms in a non-relativistic
expansion of Eq.~(\ref{eq:rpg}) reduce to the
familiar expressions for the
$\rho\pi\gamma$ charge and current operators, as
given, for example, in Ref.~\cite{Schiavilla91}.
However, it is known that these lowest-order
expansions, particularly that for ${\bf j}_{\rho\pi\gamma}$,
are inaccurate.  This fact was first demonstrated
by Hummel and Tjon~\cite{Hummel89} in the context
of relativistic boson-exchange-model calculations
of the deuteron form factors, based on the
Blankenbecler-Sugar reduction of the Bethe-Salpeter
equation.  It was later confirmed, in a
calculation of the deuteron $B(q)$ structure
function~\cite{Schiavilla91}, that next-to-leading-order
terms in the expansion for ${\bf j}_{\rho\pi\gamma}$, proportional  
$(1+\kappa_{\rho NN})/m^2$, very substantially 
reduce the contribution of the leading term.
This issue will be returned to later in Sec.~\ref{sec:res}.
Here, again we stress that the full Lorentz structure
of $j^\sigma_{\rho\pi\gamma}$ is retained in the
present study.
\section{Deuteron Form Factors}
\label{sec:ffs}

The deuteron structure functions $A$ and $B$, and tensor
polarization $T_{20}$ are expressed in terms of the
charge ($G_0$ and $G_2$) and magnetic ($G_1$) form factors as

\begin{eqnarray}
A(q)&=& G_0^2(q) + \frac{2}{3} \eta\, G_1^2(q) + \frac{8}{9} \eta^2 G_2^2(q) \>\>, \\
B(q)&=& \frac{4}{3} \eta (1+\eta) G_1^2(q) \>\>, \\
T_{20}(q)&=&-\sqrt{2} \frac{ x(x+2)+y/2}{1+2( x^2+y)} \>\>,
\end{eqnarray}
where $\eta$=$(q/2 m_d)^2$, $x$=$(2/3) \eta\, G_2(q)/G_0(q)$,
$y$=$(2/3) \eta \left[ 1/2 + (1+\eta)\, {\rm tg}^2\theta/2 \right]
\left[G_1(q)/G_0(q)\right]^2$, 
and $\theta$ is the electron scattering angle.  The
expressions above are in the Breit frame, and therefore $q$ denotes the
magnitude of the three-momentum transfer.  The form factors are
normalized as

\begin{equation}
G_0(0) = 1 \>\>,\quad G_1(0)=(m_d/m) \mu_d\>\>,\quad G_2(0)=m_d^2\, Q_d \>\>,
\end{equation}
where $\mu_d$ and $Q_d$ are the
deuteron magnetic moment (in units of n.m.) and quadrupole
moment, respectively, and are related to Breit-frame
matrix elements of the nuclear electromagnetic charge,
$\rho({\bf q})$, and current, ${\bf j}({\bf q})$, operators via:

\begin{eqnarray}
G_0(q)&=&\frac{1}{3} \sum_{M=\pm 1,0}
 \langle \psi_M;{\bf v}_B \mid \rho({\bf q}) \mid \psi_M;-{\bf v}_B \rangle
 \>\>, \label{eq:g0} \\
G_1(q)&=&-\frac{1}{\sqrt{\eta}} \langle \psi_{M=1};{\bf v}_B
\mid j_{\lambda=1}({\bf q}) \mid \psi_{M=0};-{\bf v}_B \rangle
\>\>, \label{eq:g1} \\
G_2(q)&=&\frac{1}{2\eta}\left[
 \langle \psi_{M=0};{\bf v}_B \mid \rho({\bf q}) \mid \psi_{M=0};-{\bf v}_B \rangle
-\langle \psi_{M=1};{\bf v}_B \mid \rho({\bf q})
\mid \psi_{M=1};-{\bf v}_B \rangle\right] \>\>. \label{eq:g2}
\end{eqnarray}
Here $j_{\lambda=1}$ denotes the
standard +1 spherical component of the current operator.  

The calculations are carried out in momentum space.  The
wave functions $\psi_M({\bf p};\pm {\bf v}_B)$ are given
in the second line of Eq.~(\ref{eq:boost}), and the charge and
current operators are those described in Sec.~\ref{sec:current}.
The one-body current matrix elements involve the evaluation
of integrals of the type, in a schematic notation,

\begin{equation}
\int\frac{{\rm d}{\bf p}}{(2\pi)^3}
\psi_{M^\prime}({\bf p}+{\bf q}/4; {\bf v}_B) \,
j^\sigma_{\rm 1-body}({\bf p}^\prime_1,{\bf p}_1) \,
\psi_M({\bf p}-{\bf q}/4;-{\bf v}_B) \>\>,
\end{equation}
with ${\bf p}_1^\prime$=${\bf p}+{\bf q}/2$
and ${\bf p}_1$=${\bf p}-{\bf q}/2$, which are
performed by standard Gaussian integrations.  The
two-body current matrix elements, instead, require
integrations of the type

\begin{equation}
\int\frac{{\rm d}{\bf p}^\prime}{(2\pi)^3} \frac{{\rm d}{\bf p}}{(2\pi)^3}
\psi_{M^\prime}({\bf p}^\prime; {\bf v}_B) \,
j^\sigma_{\rm 2-body}({\bf k}_1,{\bf k}_2) \,
\psi_M({\bf p};-{\bf v}_B) \>\>,
\end{equation}
with ${\bf k}_1$=${\bf q}/2+{\bf p}^\prime-{\bf p}$
and ${\bf k}_2$=${\bf q}/2-{\bf p}^\prime+{\bf p}$.
These integrations are efficiently done by Monte Carlo techniques
by sampling configurations $({\bf p},{\bf p}^\prime)$
according to the Metropolis algorithm
with a probability density
$W({\bf p},{\bf p}^\prime)$=$w({\bf p})$$w({\bf p}^\prime)$,
where $w({\bf p})$=$p^2 \mid \psi_{M=0}({\bf p};0)\mid^2$.
The computer programs have been successfully tested
by comparing, in a model calculation which ignored
boost corrections and kept only the leading terms in the
expansions for $j^\sigma_{\rm 1-body}$ and
$j^\sigma_{\rm 2-body}$, the present results with
those obtained with an earlier, configuration-space version
of the code~\cite{Schiavilla91}.
\section{Further Results}
\label{sec:res}

\begin{figure}[bth]
\let\picnaturalsize=N
\def\picsize{4in}
\def\picfilenamea{b_nr-r_jun.eps}
\ifx\nopictures Y\else{\ifx\epsfloaded Y\else\input epsf \fi
\let\epsfloaded=Y
\centerline{
\ifx\picnaturalsize N\epsfxsize \picsize\fi \epsfbox{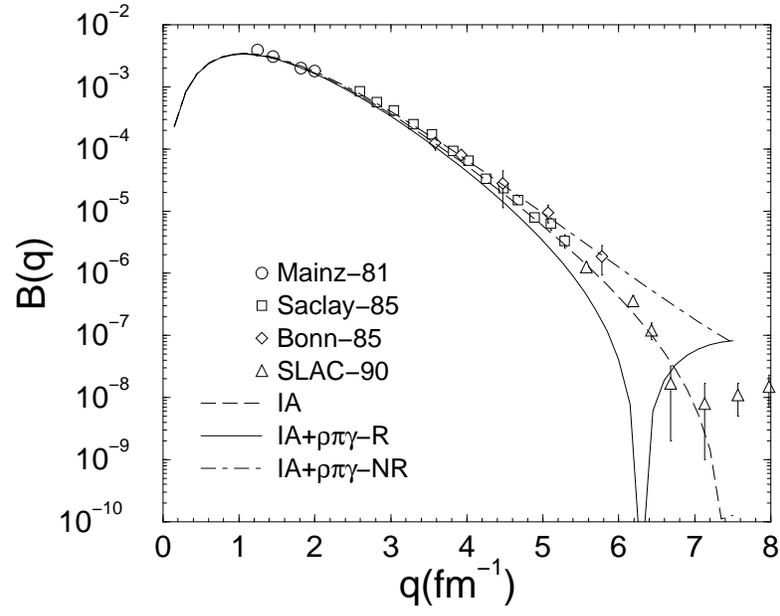}
 }}\fi
\caption{The $B(q)$ structure functions obtained with the
relativistic and non-relativistic forms of the
$\rho\pi\gamma$ current of Eqs.~(\protect\ref{eq:rpg})
and~(\protect\ref{eq:rpg_nr}), labelled respectively $\rho\pi\gamma$-R
and $\rho\pi\gamma$-NR.}
\label{fig:b_nr}
\end{figure}

In this section we briefly discuss the contribution
of the $\rho\pi\gamma$ current to the $B(q)$ structure
function.  In Fig.~\ref{fig:b_nr} the results calculated
with the current given in Eq.~(\ref{eq:rpg}) (curve
labelled $\rho\pi\gamma$-R) are compared with those
obtained by using the leading term in its non-relativistic
expansion (curve labelled $\rho\pi\gamma$-NR),

\begin{eqnarray}
{\bf j}^{\rm NR}_{\rho\pi\gamma}({\bf q})=&-&{\rm i} 
g_{\rho\pi\gamma} \frac{ g_{\rho NN} }{m_\rho} \frac{f_{\pi NN} }{m_\pi}
G_{\rho\pi\gamma}(q) 
\frac{f_\rho(k_1)}{k_1^2+m_\rho^2}
\frac{f_\pi(k_2)}{k_2^2+m_\pi^2}
({\bf k}_1 \times {\bf k}_2)\,  {\bbox \sigma}_2 \cdot {\bf k}_2
\, {\bbox \tau}_1 \cdot {\bbox \tau}_2 \nonumber \\
&+& 1 \rightleftharpoons 2 \ .
\label{eq:rpg_nr}
\end{eqnarray}
Figure~\ref{fig:b_nr} demonstrates the inadequacy of the
approximation~(\ref{eq:rpg_nr}), a fact which, as mentioned already
in Sec.~\ref{sec:current}, has been known for some time~\cite{Hummel89}.
Indeed a more careful analysis shows that the contributions of
next-to-leading order terms are not obviously negligible, since they are
proportional to the large $\rho$$N$$N$ tensor coupling constant,
$\kappa_{\rho NN}$=$6.1$ in Ref.~\cite{Machleidt01}.  We sketch
the derivation again here for the sake of completeness.

The vector structure $\epsilon_{i\nu\sigma\tau} \Gamma^\nu \, k_1^\sigma \, q^\tau$
occurring in Eq.~(\ref{eq:rpg}), with $\Gamma^\nu=(\Gamma^0,\bbox{\Gamma})$
defined as

\begin{equation}
\Gamma^\nu \equiv \overline{u}_1^\prime
\left( \gamma^\nu +{\rm i} \frac{\kappa_{\rho NN} }{2 m}
\sigma^{\nu \alpha} k_{1\,\alpha} \right) u_1 \ ,
\end{equation}
can be written in the Breit frame as

\begin{equation}
\epsilon_{i\nu\sigma\tau} \Gamma^\nu \, k_1^\sigma \, q^\tau =
\Gamma^0\, \left({\bf q} \times {\bf k}_1\right)
-k_1^0\, \left({\bf q}\times \bbox{\Gamma}\right) \ ,
\end{equation}
where the index $i=1,2,3$ and ${\bf q}$=${\bf k}_1+{\bf k}_2$.
Up to order $(v/c)^2$ included, one finds, in analogy
to the non-relativistic expansion of the nucleon
electromagnetic current (with point-nucleon couplings,
i.e. $F_1^S(q)\rightarrow 1$ and $F_2^S(q)\rightarrow \kappa_{\rho NN}$)

\begin{eqnarray}
\Gamma^0&\simeq& 1-\left(1+2 \kappa_{\rho NN}\right)\left[\frac{k_1^2}{8 m^2}
-\frac{\rm i}{4 m^2} \bbox{\sigma}_1 \cdot
\left({\bf p}_1^\prime \times {\bf p}_1\right)\right]\ ,\label{eq:rpgv1} \\
\bbox{\Gamma}&\simeq&\frac{1}{2 m}\left({\bf p}_1^\prime +{\bf p}_1\right)
+\frac{\rm i}{2 m} \left(1+\kappa_{\rho NN}\right)
\bbox{\sigma}_1 \times {\bf k}_1 \ .
\label{eq:rpgv2}
\end{eqnarray}
Retaining only the leading term in $\Gamma^\nu$,
namely $\Gamma^\nu\simeq (1,0)$, leads
to the expression in Eq.~(\ref{eq:rpg_nr}).
Some of the corrections proportional to $\kappa_{\rho NN}$ were
explicitly calculated in Ref.~\cite{Schiavilla91}, and
were found to decrease substantially those from the leading term.

We note, in passing, that the $\rho\pi\gamma$ charge operator,
$\rho_{\rho\pi\gamma}({\bf q})$, is proportional to

\begin{eqnarray}
\epsilon_{i\nu\sigma\tau} \Gamma^\nu \, k_1^\sigma \, q^\tau&=&
\bbox{\Gamma} \cdot {\bf k}_1 \times {\bf q} \nonumber \\ 
&\simeq& \frac{\rm i}{2 m}\left( 1+\kappa_{\rho NN}\right) \bbox{\sigma}_1
\cdot {\bf k}_1 \times {\bf k}_2 \ ,
\end{eqnarray}
where the small non-local term proportional
to ${\bf p}_1^\prime+{\bf p}_1$ has been neglected.  The
standard form of the $\rho_{\rho\pi\gamma}({\bf q})$ commonly
used in studies of light nuclei form factors~\cite{Marcucci98,Wiringa98}
easily follows.  In this case too, however, higher order
corrections included in $j^0_{\rho\pi\gamma}$, Eq.~(\ref{eq:rpg}),
reduce the contribution of the leading term, although they
do not change its sign.  A more detailed discussion of this
issue will be given in Ref.~\cite{Schiavilla02}. 

We conclude this section with a couple of remarks. The first
is that the destructive interference between the one-body and
$\rho\pi\gamma$ currents is also obtained in
recent calculations of the $B(q)$ structure function, carried
out in the covariant framework based
on the spectator equation~\cite{VanOrden95,Gilman02}.

The second remark is that in all earlier studies of light
nuclei form factors~\cite{Carlson98} additional
two-body currents, originating from the
momentum-dependent terms of the two-nucleon potential,
were included.  Work is in progress~\cite{Schiavilla02}
on an improved treatment of these currents, in an approach 
similar to that proposed in Ref.~\cite{Tsushima93}.
\section*{Acknowledgments}
The work of R.S. was supported by the U.S. Department 
of Energy contract DE-AC05-84ER40150, under which the
Southeastern Universities Research Association (SURA)
operates the Thomas Jefferson National Accelerator Facility.
The work of V.R.P. was supported by the U.S. National
Science Foundation via grant PHY 00-98353.  Finally,
most of the calculations were made possible by grants
of computing time from the National Energy Research
Supercomputer Center.
\end{document}